\input harvmac

\def \s {\sigma}

\def \ha {\half}
\def \ov {\over}

\def \a {\alpha}
\def \lr { \lref}

\def \del {\partial}

\def \ha{{\textstyle{1\over 2}}}

\def \a {\alpha}

\def \zeta {\zeta}
\def \s {\sigma}

\def \td {\tilde }

\def \ov {\over }

\def   \td {\tilde }

\def\({\left( }
\def\){\right)}
\def\N{ {\cal N} }

\def\tre{ {\textstyle{1\ov 3} } }


\def \lr { \lref}
\def\np {{  Nucl. Phys. }}

\lref\thooft{G. 't Hooft, ``A Planar Diagram Theory for Strong Interactions,'' Nucl. Phys. {B72} (1974) 461.}

\lref\polya{A. M. Polyakov, ``String Theory and Quark Confinement,''
hep-th/9711002.}

\lref\malda{J. Maldacena,  ``The Large
$N$ Limit of Superconformal Field Theories and Supergravity,''
hep-th/9711200.} 

\lref\ewt{E. Witten, ``Some Comments on String Dynamics,'' in {\it Strings '95},
ed. I. Bars et. al. (World Scientific, 1997), hep-th/9507121;
N.~Seiberg and E.~Witten, \np B471 (1996) 121.
}

\lr\hoog{G. T. Horowitz and H. Ooguri, ``Spectrum of Large $N$ Gauge Theory
from Supergravity,'' hep-th/9802116.}

\lref\oldstr{A. Strominger, ``Open $p$-Branes,'' Phys. Lett. {\bf B383} (1996) 44, hep-th/9512059.} 

\lr\horstro{G.~T. Horowitz and A.~Strominger, \np B360 (1991) 197.}

\lref\duff{M. J. Duff, B. E. W. Nilsson,
and C. N. Pope, ``Kaluza-Klein Supergravity,'' Physics Reports
{\bf 130} (1986) 1.}

\lref\zys{R.~de~Mello~Kock, A.~Jevicki, M.~Mihailescu and J.~Nunes, ``Evaluation of glueball masses from supergravity", hep-th/9806125;
M.~Zyskin, ``A note on the glueball mass spectrum", hep-th/9806128;
J.~Greensite and P.~Olesen, ``Remarks on the heavy quark potential in the supergravity approach",
hep-th/9806235.}

\lref\oduff{M. J. Duff, H. Lu,
and C. N. Pope, ``$AdS_5\times S^5$ Untwisted,'' hep-th/9803061.} 

\lr\kach{S. Kachru and E. Silverstein, ``4d Conformal Field Theories
and Strings on Orbifolds,'' hep-th/9802183.}

\lr\vaffa{M. Bershadsky, Z. Kakushadze, and C. Vafa,
``String Expansion as Large $N$ Expansion of Gauge Theories", hep-th/9803076.}

\lr\edw{E. Witten,  ``Anti-de Sitter Space, Thermal Phase Transition, and Confinement in Gauge Theories'',
hep-th/9803131.}

\lr\wittn{E. Witten,  ``Anti-de Sitter Space and Holography,''
hep-th/9802150.}

\lref\mastro{J.~Maldacena and A.~Strominger, ``$AdS_3$
Black Holes and a Stringy Exclusion Principle",
hep-th/9804085.}

\lr\gkp{S.~S. Gubser, I. R. Klebanov, and A. M. Polyakov, ``Gauge Theory Correlators from Noncritical String Theory,'' hep-th/9802109.}

\lr\russo{J.G. Russo, ``Einstein spaces in five and seven dimensions and non-supersymmetric
gauge theories", Phys.Lett.~B435 (1998) 284, hep-th/9804209.}

\lr\klebw{I.~Klebanov and E.~Witten, ``Superconformal field theory on threebranes at a Calabi-Yau singularity", hep-th/9807080.}

\lr\keha{A.~Kehagias, ``New type IIB vacua and their F-theory interpretation", hep-th/9805131.}

\lr\gubser{S.~S. Gubser, ``Einstein manifolds and conformal field theories",
hep-th/9807164.}

\lr\cvet{M. Cveti\v c and D. Youm, ``Rotating intersecting M-branes",
\np B499 (1997) 253.}

\lr\horsen{G.~Horowitz and A.~Sen, ``Rotating black holes which saturate a 
Bogomol'nyi bound",  Phys.Rev. D53 (1996) 808.}

\lr\groo{D.J.~Gross and H.~Ooguri, ``Aspects of large $N$ gauge theory dynamics as seen by string theory", hep-th/9805129.}

\lr\oogu{C.~Cs\' aki, H.~Ooguri, Y.~Oz and J.~Terning,
``Glueball mass spectrum from supergravity", hep-th/9806021.}

\lr\ogkk{H.~Ooguri, H.~Robins and J.~Tannenhauser, ``Glueballs and their Kaluza-Klein cousins", hep-th/9806171.}

\lr\itza{A.~Brandhuber, N.~Itzhaki, J.~Sonnenschein and S.~Yankielowicz,
``Wilson loops, confinement and phase transitions in large $N$ gauge theories from supergravity", hep-th/9803263.}

\lr\itzma{N.~Itzhaki, J.~Maldacena, J.~Sonnenschein and S.~Yankielowicz, ``Supergravity and the large $N$ limit of theories with sixteen supercharges",
hep-th/9802042.}

\lr\mwils{J.~Maldacena, ``Wilson loops in large $N$ gauge theories",
Phys. Rev. Lett. 80 (1998) 4859,
hep-th/9803002.}

\lr\aha{O.~Aharony, A.~Fayyazuddin and J.~Maldacena, ``The large $N$ limit of $\N=2,1$ field theories from threebranes in $F$-theory",
hep-th/9806159.}




\baselineskip8pt
\Title{
\vbox
{\baselineskip 6pt{\hbox{ }}{\hbox
{Imperial/TP/97-98/68}}{\hbox{hep-th/9808117}} {\hbox{
  }}} }
{\vbox{\centerline {New Compactifications of Supergravities}
\vskip4pt
 \centerline      {and Large $N$ QCD}
}}
\vskip -27 true pt
\bigskip
\bigskip

\centerline  {  Jorge G. Russo{\footnote {$^*$} {e-mail address:
j.russo@ic.ac.uk
 } } 
}

\smallskip
\bigskip

 \centerline {\it  Theoretical Physics Group, Blackett Laboratory,}
\smallskip
\centerline {\it  Imperial College,  London SW7 2BZ, U.K. }

\medskip\bigskip

\centerline {\bf Abstract}
\medskip
\baselineskip14pt
\noindent

We construct  supergravity backgrounds representing non-homogeneous 
compactifications of $d=10,11$ supergravities to four dimensions, which cannot be written as a direct product. The geometries are regular and
approach $AdS_7\times S^4$ or $AdS_5\times S^5$ at infinity; they
are generically non-supersymmetric, except in a certain ``extremal" limit, where a Bogomol'nyi bound is saturated and
a naked singularity appears.
By using these spaces, one can construct
 a model of QCD that 
generalizes  by one (or two) extra parameters a recently proposed model of QCD based on the non-extremal D4 brane.
This allows for some extra freedom to circumvent some (but not all) 
limitations of the simplest version.

\medskip
\Date {August 1998}
\noblackbox
\baselineskip 17pt plus 2pt minus 2pt

\newsec{Introduction }

In the last few months there have been a number of suggestions
generalizing the dualities proposed by Maldacena \malda\ 
to the case of four-dimensional gauge theories with less supersymmetries (see e.g. refs.~\refs{\wittn \kach  \vaffa \edw \russo\keha \aha \klebw -\gubser}). 
In most of the examples so far considered, the supergravity backgrounds are of the form $AdS_n\times X$, where $X$ is some Einstein manifold, and they are 
expected to be related to a conformal field theory ``at the boundary" 
\refs{\malda, \gkp,\wittn }, but, because of conformal invariance, not to a confining gauge theory (for a review of supergravity compactifications on anti-de Sitter backgrounds, see \duff ).
The interesting model introduced by Witten \edw\ and further investigated in
\refs{\groo \oogu \zys -\ogkk } exhibits confinement, but, as pointed out
in these works, its applications to $3+1$ dimensional QCD are somewhat 
limited by the fact that there is a single scale in the geometry.
The model arises as dimensional reduction of $4+1$ dimensional $SU(N)$ super Yang-Mills theory, which as quantum field theory is non-renormalizable;
in order to decouple Kaluza-Klein states, 
at least two scales seem necessary: one representing $\Lambda_{\rm QCD}$, and another one for the masses of Kaluza-Klein particles $M_{\rm KK}$, with $\Lambda_{\rm QCD}\ll M_{\rm KK}$.
In addition, understanding certain properties of the spectrum seems to  require an understanding of string theory in Ramond-Ramond backgrounds. 
Unfortunately, such string models are not solvable; 
 to be able to  uncover the properties of large $N$ QCD, one would like to have
a regime where there is a supergravity description.

The present work  is an attempt in this direction. 
We will assume that a supergravity description of pure 
Yang-Mills theory at large 't Hooft coupling exists and start with
the most general regular geometry that  can plausibly describe a confining theory, in the sense that will be explained below.
The non-supersymmetric background of section 3 generalizes
the Witten QCD model  in much the same way the Kerr black hole  generalizes the Schwarzschild black hole solution. It is, however, a static geometry 
--~what is ``time" in the Kerr solution here plays the role of an angular coordinate~--, which
is regular  everywhere, and contains  an extra parameter $a$ with respect to the Witten QCD model.
When this parameter is small, the model reduces to the Witten model, with 
$\Lambda_{\rm QCD}\cong M_{\rm KK}$.
When this parameter is large, one obtains the desired decoupling
of unwanted Kaluza-Klein particles, $\Lambda_{\rm QCD}\ll M_{\rm KK}$,
which guarantees that the gauge theory at the QCD scale is a $3+1$ (rather than $4+1$) dimensional one.

\newsec{Large $N$ QCD from supergravity}

In order to introduce our notation, in this section we review the approach to 
non-supersymmetric Yang-Mills theory of \edw .

\subsec{Construction of the metric from the black M5-brane}

We start with the metric for the
non-extremal M5-brane of eleven-dimensional supergravity, which is given by
\eqn\buno{
ds^2_{11}=f^{-1/3}\big[ -h \ dt^2+dx_1^2+...+dx^2_5\big]+
f^{2/3}\bigg[{dr^2\ov  h} +r^2 d\Omega_4^2\bigg]\ ,
}
where
\eqn\bdos{
f=1+{2m \sinh^2\a\ov r^3}\ ,\ \  \ h=1-{2m\ov r^3  }\ ,\ \ 
}
There is in addition non-zero flux of the four-form field strength in the four-sphere with
quantized (magnetic) charge $N$ related to $m$ and $\a $ by
\eqn\btres{
2m \cosh\a \sinh\a =\pi N l_P^3\ ,
}
where $l_P$ is the Planck lenght in eleven dimensions (in terms of string theory parameters, defined as $l_P=g^{1/3}\sqrt{\a' }$).
There is an event horizon at $r=r_H$, $r_H=(2m)^{1/3}$. 
The Hawking temperature is given by 
\eqn\bcinco{
T_H={3r_H^2\ov 8\pi m\cosh\a } \ .
}
By dimensional reduction in $x_5$, one obtains the type IIA solution representing the
non-extremal D4-brane \horstro :
\eqn\bsiete{
ds^2_{\rm IIA}=f^{-1/2}\big[ -h \ dt^2+dx_1^2+...+dx^2_4\big]+
f^{1/2}\bigg[{dr^2\ov  h} +r^2d\Omega_4^2\bigg] \ ,
}
\eqn\sie{
e^{2(\phi(r)-\phi_{\infty})}=f^{-1/2}(r)\ .
}
The supergravity solution that is related to a field theory is obtained by taking the (``decoupling") low-energy limit $l_P\to 0$ in \buno . 
This is done by rescaling variables as follows~\refs{\malda, \itzma}:
\eqn\deco{
r=U^2 l_P^3\ ,\ \ \ \ m=\ha U_0^6 l_P^9\ ,\ 
}
and taking the limit $l_P\to 0$ at fixed $U,U_0$, so that
$2m\sinh^2\a \to \pi Nl_P^3$ (see \btres ).
We get
\eqn\once{
ds^2_{11}=l_P^2 \bigg[ {U^2\ov (\pi N)^{1/3} }
\big[ -(1-{U_0^6\ov U^6 } ) dt^2+dx_1^2+...+dx_5^2\big]+
{4(\pi N)^{2/3}dU^2\ov U^2(1-{U_0^6\ov U^6})}
+ (\pi N)^{2/3}d\Omega_4^2\bigg]\ .
}
There is a horizon at $U=U_0$.
At infinity, the solution approaches the geometry of $AdS_7\times S^4$.
It can also be obtained as a special limit of the Schwarzschild-de Sitter black 
hole~\edw .

\subsec{Witten QCD model}

The low-energy theory on the non-extremal D4 brane is
$4+1$ dimensional $SU(N)$ Yang-Mills theory at finite temperature $T_H$.
In the path integral formulation, the  gauge theory at finite temperature 
is described by going to Euclidean time $t\to -i\tau $, and identifying
$\tau $  periodically  with period $1/T_H$.
Zero-temperature $3+1$ Yang-Mills theory can be described
by making $x_4\to -i x_0$, and viewing  
$\tau $ as parametrizing a space-like circle with radius $R_0=(2\pi T_H)^{-1}$, where --~as in the thermal ensemble~--
fermions obey antiperiodic boundary conditions. At energies much lower
than $1/R_0$, 
the theory is effectively $3+1$ dimensional.
Because of the boundary conditions, fermions 
 and scalar particles get  masses proportional to the inverse radius,
so that, as $R_0\to 0$, they should decouple from the low-energy physics.
The low-energy theory is thus pure Yang-Mills theory.

The gauge coupling $g_4^2$ in the $3+1$ dimensional Yang-Mills theory is given by the ratio between the periods of the eleven-dimensional coordinate $x_5$ and
$\tau $. It is convenient to introduce ordinary angular coordinates 
$\theta_1,\ \theta_2$ which are $2\pi $-periodic by
\eqn\angg{
\tau =R_0 \theta_2\ ,\ \ \ \ x_5=g_4^2 R_0 \theta_1=
{\lambda \ov  N} R_0 \theta_1\ ,
}
\eqn\ttt{
R_0=(2\pi T_H)^{-1}= {2 (\pi N)^{1/2}\ov 3U_0 }\ ,
}
where we have introduced the 't Hooft coupling 
$$
\lambda\equiv g_4^2N\ .
$$
The eleven-dimensional metric takes the form
$$
ds^2_{11}= {U^2\ov (\pi N)^{1/3} }
\big[ -dx_0^2+dx_1^2+dx_2^2+dx^2_3\big]
+{4U^2\ov 9U_0^2}(\pi N)^{2/3} (1-{U_0^6\ov U^6 })d\theta^2_2
$$
\eqn\sonce{
+\ {4U^2\lambda^2\ov 9U_0^2N^2}(\pi N)^{2/3} d\theta^2_1
+ {4(\pi N)^{2/3}dU^2\ov U^2(1-{U_0^6\ov U^6})}
+ (\pi N)^{2/3}d\Omega_4^2\ .
}
In the large $N$ limit at fixed $\lambda $, the circle parametrized by $\theta_1$
is much smaller than that parametrized by $\theta_2$, so that we can dimensionally reduce
in $\theta_1 $.
We obtain ($U=2 (\pi N)^{1/2} u$)
$$
ds^2_{\rm IIA}={8\pi \lambda u^3\ov  3u_0} \big[ -dx_0^2+dx_1^2+dx_2^2+dx^2_3\big]+
{8\pi\lambda \ov 27}{u^3\ov u_0^3}(1-{u_0^6\ov u^6 }) d\theta_2^2
$$
\eqn\socho{
+\ {8\pi\lambda \ov 3}{du^2\ov u_0u(1-{u_0^6\ov u^6 }) }+{2\pi\lambda \ov 3u_0} ud\Omega_4^2\ \ ,
}
\eqn\dila{
e^{2\phi }={8\pi\ov 27} {\lambda^3 u^3\ov u_0^3} {1\ov N^2} \ ,
\ \ \ \ \ R_0^{-1}=3u_0 \ .
}

Wilson loops in the model exhibit a confining area-law behavior \refs{\edw,\groo , \itza}, where the string tension $T_{\rm YM}$ is proportional to
the coefficient of $\sum_{i=0}^3 dx_i^2$ at the horizon,
\eqn\sst{
T_{\rm YM}={4\ov 3 } \lambda u_0^2\ .
}

The supergravity fields can be classified in terms of representations
of the isometry group of the internal space $SO(5)\times U(1)$. Glueballs are expected to be associated with singlets of $SO(5)$ carrying vanishing $U(1)$ charge \groo .
The  squared mass of glueballs should be proportional to the string tension,
i.e.
\eqn\gll{
M_{\rm glueballs}\cong \Lambda_{\rm QCD}=u_0 \ .
}
The Kaluza-Klein states associated with the circle parametrized by $\theta_2 $ are of the form $\Phi= \chi (u) e^{ik.x} e^{in\theta_2}$.
These states will have mass of order $R_0^{-1}$, i.e. in our units,
\eqn\kkz{
M_{\rm KK}\cong R_0^{-1}=3 u_0\ .
}
Comparing eqs.~\gll\ and \kkz\ one sees that
Kaluza-Klein states cannot be decoupled
while keeping  the glueballs in the spectrum. In other words, as long
as $\Lambda_{\rm QCD}$ is finite, the theory is always $4+1$ dimensional.
In sect.~3 we explore a generalization of the Witten model which overcomes this problem.

\newsec{Generalizations}

In the holographic picture \wittn ,  $SU(N)$ gauge theory 
on ${\bf R}^4$ is expected to be
described in terms of a sum  over geometries that have ${\bf R}^4$ isometry and
asymptotically approach $AdS_7\times S^4$ (or $AdS_5\times S^5$).
A natural question is whether there are other metrics (beside the non-extremal D4 brane) that can be relevant for a description of QCD within the framework of supergravity.
We first look for regular geometries that approach $AdS_7\times S^4$ at infinity,
i.e. something of the form
$$
ds^2_{11}=U^2 h_1(U,\varphi_m) \sum_{i=0}^3 dx_i^2
+{R_{AdS}^2 dU^2\ov U^2 h_2(U,\varphi_m)} + {\rm angular \ part}\ ,
$$
where $h_1, h_2\to 1$ at $U\to \infty $. In order to have an area law
for Wilson loops within a supergravity description, it is necessary
that the coefficient of $\sum dx_i^2 $ be bounded above zero~\edw.
This suggests that there must be a horizon at some $U=U_H>0$.
No-hair theorems imply that the most general stationary  M5 metric with a regular horizon
is given by a Kerr-type metric with angular momentum components in two planes,
and they preclude any other possibility.
In the present case, since we want full Poincar\' e symmetry in the four-dimensional space $x_i$, we look for a static geometry, so 
what  plays the role of ``time" in the Kerr metric here will be an angular 
coordinate describing an internal circle (just as in the model
of the preceding section based on the non-extremal M5 brane).
Alternatively, one may  try to look for  geometries
with a regular horizon and  ${\bf R^4}$ isometry
that approach $AdS_5\times S^5$ at infinity. The no-hair theorem implies that there are none (black D3 branes do not have
${\bf R^4}$ isometry). On the face of it, there are no other 
options to describe a $3+1$ dimensional confining gauge theory other than a Kerr-type generalization of the M5 brane.
Standing possibilities are of course singular metrics, but these require an understanding of the corresponding string theory.

\subsec{A static D4-brane}
We start with the metric for the rotating M5 brane of eleven-dimensional supergravity with angular component in one plane.
This solution can be obtained by uplifting the magnetically charged rotating black hole in $D=6$ \horsen\ and it was explicitly 
constructed in  \cvet . It is given by 
$$
ds^2_{11}=f^{-1/3}\big[ -h \ dt^2+dx_1^2+...+dx^2_5\big]+
f^{2/3}\bigg[{dr^2\ov \td h} +r^2\big( \Delta d\theta^2+\td \Delta \sin^2\theta d\varphi^2 
$$
\eqn\uno{
+\ \cos^2\theta\ d\Omega_2^2\big)
 -{4lm\cosh \a \ov r^3\Delta f} \sin^2\theta dtd\varphi \bigg]\ ,
}
where
\eqn\dos{
f=1+{2m \sinh^2\a\ov r^3\Delta }\ ,\ \ \ \Delta=1+{l^2\cos^2\theta\ov r^2}\ ,
\ \ \ \td \Delta=1+{l^2\ov r^2} +{2ml^2 \sin^2\theta\ov r^5 \Delta f}\ ,
}
\eqn\tres{
h=1-{2m\ov r^3 \Delta }\ ,\ \ \ \ \td h={1\ov \Delta}\big(1+{l^2\ov r^2} 
-{2m\ov r^3}\big)\ ,
}
$$
d\Omega_2^2= d\psi_1^2+\sin^2\psi_1 d\psi_2^2 \ .
$$
There are in addition non-zero components for the three-form field
$$
C_{\varphi\psi_1\psi_2}=2m \Delta^{-1} (1+{l^2\ov r^2}) \cosh\a \sinh\a
\cos^2\theta\ ,\ \ \ \ \ C_{t\psi_1\psi_2}=-{2ml\ov r^2\Delta }\sinh\a
\cos^2\theta \ .
$$
The quantized (magnetic) charge $N$ is related to $\a $ and $m$ as
in \btres .
There is an event horizon at $r=r_H$, where $r_H$ is the real solution of
$r^3+l^2r-2m =0$, which exists for all values of $m>0$ and $l$. 
The Killing vector $\del/\del t$ diverges at $r=0$ and becomes space-like in the ergosphere region outside the horizon where $r^2+l^2\cos^2\theta -2m/r <0$.
There is no inner horizon, just a space-like singularity at $r=0$. In the limit $m\to 0$, horizon and singularity coalesce, and
the solution saturates the BPS bound \horsen .
The Hawking temperature is given by
\eqn\cinco{
T_H={3r_H^2+l^2\ov 8\pi m\cosh\a }\ .
}
By dimensional reduction in $x_5$, one obtains the following type IIA solution, representing a rotating black D4-brane:
$$
ds^2_{\rm IIA}=f^{-1/2}\big[ -h \ dt^2+dx_1^2+...+dx^2_4\big]+
f^{1/2}\bigg[{dr^2\ov \td h} +r^2\big( \Delta d\theta^2+\td \Delta \sin^2\theta d\varphi^2 
$$
\eqn\seis{
+ \ \cos^2\theta \ d\Omega_2^2\big)
 -{4lm\cosh \a \ov r^3\Delta f} \sin^2\theta dtd\varphi\bigg] \ ,
}
\eqn\siete{
e^{2(\phi (r) -\phi_{\infty})}=f^{-1/2} \ .
}
Although string theory on this space cannot be dual to Lorentz-invariant quantum field theory,
a metric with ${\bf R^4} $ Poincar\' e symmetry can nevertheless be obtained by the formal substitution:
$$
t\to -i\tau\ ,\ \ \ \ x_4\to ix_0\ ,\ \ \ \ l\to ib
$$
where $\tau $ is periodic of period $1/T_H$. We obtain
$$
ds^2_{\rm IIA}=f^{-1/2}\big[ -dx_0^2+dx_1^2+dx_2^2+dx^2_3\big]+f^{-1/2} h\ d\tau^2
+f^{1/2}\bigg[{dr^2\ov \td h} 
$$
\eqn\ocho{
+\ r^2\big( \Delta d\theta^2+\td \Delta \sin^2\theta d\varphi^2
+ \cos^2\theta d\Omega_2^2\big)
 -{4b m\cosh \a \ov r^3\Delta f} \sin^2\theta d\tau d\varphi \bigg] \ ,
}
with $\phi $ and $\Delta, \td \Delta, \td h$ as before with $l\to ib $, i.e.
\eqn\nueve{
\Delta=1-{b^2\cos^2\theta\ov r^2}\ ,
\ \ \ \td \Delta=1-{b^2\ov r^2} -{2mb^2 \sin^2\theta\ov r^5 \Delta f}\ ,
\ \ \ \td h={1\ov \Delta}\big(1-{b^2\ov r^2} -{2m\ov r^3}\big)\ .
}
The geometry is regular and geodesically complete (being essentially the Euclidean Kerr metric) with the space-time restricted to the region $r>r_H$, where  $r_H$ is 
the real solution of $r^3-b^2r-2m=0$. 
Since $m>0$, one has $r_H>b$ (there is a singularity at the surface $r^2=b^2\cos^2\theta $, but this lies inside the horizon).
It should be noted that the interpretation as rotating D4-brane is no longer valid. The geometry now represents  a static 4-brane configuration with Ramond-Ramond charge,
where one of the world-volume directions $\tau $ is compactified on a circle and mixed with  the angular directions.

Let us now consider the eleven-dimensional metric that is obtained in the ``decoupling" limit. As in \deco , we redefine variables as follows:
\eqn\diez{
r=U^2 l_P^3\ ,\ \ \ \ m=\ha U_0^6 l_P^9\ ,\ \ \ \ b=a^2l_P^3\ ,
}
and take the limit $l_P\to 0$ at fixed $U,U_0$ and $a$. 
We get
$$
ds^2_{11}=l_P^2 \Delta^{1/3}\bigg[ {U^2\ov (\pi N)^{1/3} }
\bigg( -dx_0^2+dx_1^2+dx_2^2+dx^2_3+dx_5^2+\big(
1-{U_0^6\ov U^6\Delta }\big)  d\tau^2
\bigg)+{4(\pi N)^{2/3}dU^2\ov U^2(1-{a^4\ov U^4}-{U_0^6\ov U^6})}
$$
\eqn\once{
+\ (\pi N)^{2/3}\bigg(d\theta^2+{\td \Delta\ov \Delta} \sin^2\theta d\varphi^2 
+{1\ov \Delta } \cos^2\theta d\Omega_2^2
 -{2a^2 U_0^3\ov U^4\Delta (\pi N)^{1/2} } \sin^2\theta d\tau d\varphi
\bigg)\bigg]\ ,
}
where
$$
\Delta=1-{a^4\cos^2\theta \ov U^4}\ ,\ \ \ \ \td \Delta=1-{a^4\ov U^4} \ .
$$
The background thus represents a smooth compactification of 
$d=11$ supergravity which is not of ``direct product" form. Near infinity, it looks like
$AdS_7\times S^4$. 

Similarly, one can construct the more general solution with parameters
$\{ U_0, N,a_1,a_2 \}$ by starting with the M5-brane with angular momentum in two planes \cvet , with parameters $\{ m, N, l_1,l_2\} $, and making
$t\to -i\tau $, $l_{1,2}=ia_{1,2}^2 l_P^3,\ 2m=U_0^6l_P^9$.
In the decoupling limit, it is of the form
$$
ds^2_{11}=l_P^2 \bigg[ {U^2\ov (\pi N)^{1/3} }\Delta^{1/3}
\bigg( -dx_0^2+dx_1^2+dx_2^2+dx^2_3+dx_5^2+\big(
1-{U_0^6\ov U^6\Delta }\big)  d\tau^2
\bigg)
$$
\eqn\docc{
+\ {4(\pi N)^{2/3}\Delta^{1/3}dU^2\ov U^2
(1-{a^4_1\ov U^4}-{a^4_2\ov U^4} + {a^4_1a^4_2\ov U^8}-{U_0^6\ov U^6})}
+\ {\rm angular\ part}\bigg]\ ,
}
\eqn\abc{
\Delta= 1-{a_1^4\ov U^4} \cos^2\theta -
{a_2^4\ov U^4}\big( \cos^2\theta \cos^2\psi_1 +\sin^2\theta\big)
+{a^4_1a^4_2\ov U^8}\cos^2\theta \cos^2\psi_1 \ .
}

\subsec{Type IIB vacua}

Analogous spaces containing 3-branes of type IIB theory, which are of 
the form $AdS_5\times S^5$ at infinity, can be constructed by starting with a 
 ``stack" of rotating M2 branes in eleven dimensions with two extra translational isometries.
This is obtained by a slight generalization of the solutions of \refs{\horsen, \cvet}. The metric is given by
$$
ds^2_{11}=f^{-2/3}_0\big[ -h_0 \ dx_0^2+dx_1^2+dx^2_2\big]+
f_0^{1/3}\bigg[{dr^2\ov \td h_0} +d\tau_1^2+d\tau_2^2+
r^2\big( \Delta_0 d\theta^2+\td \Delta_0 \sin^2\theta d\varphi^2 
$$
\eqn\doce{
+\ \cos^2\theta \ d\Omega_3^2\big)
 -{4lm\cosh \a \ov r^4\Delta_0 f_0} \sin^2\theta\ dx_0d\varphi 
\bigg]\ .
}
The functions $f_0,\Delta_0, $ etc. are as before with the replacement
$m/r^3\to m/r^4$, that is,
\eqn\trece{
f_0=1+{2m \sinh^2\a\ov r^4\Delta_0 }\ ,\ \ \ \Delta_0=1+{l^2\cos^2\theta\ov r^2}\ ,
\ \ \ \td \Delta_0=1+{l^2\ov r^2} +{2ml^2 \sin^2\theta\ov r^6 \Delta_0 f_0}\ ,
}
\eqn\auno{
h_0=1-{2m\ov r^4 \Delta_0 }\ ,\ \ \ \ \td h_0={1\ov \Delta_0}
\big(1+{l^2\ov r^2} 
-{2m\ov r^4}\big)\ .
}
Upon dimensional reduction in $\tau_2$ and T-duality in $\tau_1$ 
one finds
$$
ds^2_{\rm IIB}=f^{-1/2}_0\big[ -h_0 \ dx_0^2+dx_1^2+dx^2_2+dx_3^2\big]+
f_0^{1/2}\bigg[{dr^2\ov \td h_0} +
r^2\big( \Delta_0 d\theta^2+\td \Delta_0 \sin^2\theta d\varphi^2 
$$
\eqn\ados{
+\ \cos^2\theta \ d\Omega_3^2\big)
 -{4lm\cosh \a \ov r^4\Delta_0 f_0} \sin^2\theta \ dx_0d\varphi \bigg]\ ,
}
$$
e^{\phi_B}=g_s={\rm const.}
$$
where we have replaced the compact coordinate T-dual to $\tau_1$ by an
uncompact coordinate $x_3$. 
The solution \ados\ represents a rotating black D3-brane with
charge $N$, related to $m,\a $ by
\eqn\queq{
4\pi g_sN{\a'}^2=2m\cosh\a\sinh\a\ .
}
There is an event horizon at $r^4+l^2r^2-2m=0$, or
$r_H^2=\ha (\sqrt{l^4+8m}-l^2)$. For any $m>0$, there is a space-like
singularity at $r=0$. The ADM mass per unit volume of the three brane
is given by
$$
{M_{\rm ADM}\ov V_3}= {c m\ov 2\pi g_s {\a'}^2 }  \big(\cosh^2\a +
{1\over 4} \big)\ .
$$

Let us now consider the extremal limit $m\to 0$ at fixed charge $N$.
In this limit the solution saturates the Bogomol'nyi bound ${M_{\rm ADM}\ov V_3}=c N$
(there should be 16 unbroken supersymmetries, i.e. 1/2 of the supersymmetries
of $AdS_5\times S^5$).  
We obtain
$$
ds^2_{\rm IIB}=f^{-1/2}_0\big[ - dx_0^2+dx_1^2+dx^2_2+dx_3^2\big]+
f_0^{1/2}\bigg[{(r^2+l^2 \cos^2\theta)dr^2\ov r^2+l^2} 
$$
\eqn\atres{+\ 
(r^2+l^2 \cos^2\theta)d\theta^2+(r^2+l^2) \sin^2\theta d\varphi^2 
+ r^2\cos^2\theta d\Omega_3^2 \bigg]\ ,
}
\eqn\qqa{
f_0=1+{4\pi g_sN{\a'}^2\ov r^4\Delta_0}\ .
}
We now rescale the variables 
$$
r=U\a'\ ,\ \ \ \ \ l=a\a'\ ,
$$
and take
the limit $\a'\to 0$, with $U,a$ fixed.
We obtain the following type IIB background:
$$
ds^2_{\rm IIB}=\a' \Delta^{1/2}_0\bigg[{U^2\ov \sqrt{4\pi g_sN} }
\big[ - dx_0^2+dx_1^2+dx^2_2+dx_3^2\big]+
\sqrt{4\pi g_sN} {dU^2\ov (U^2+a^2) }
 $$
\eqn\aatres{+\ 
\sqrt{4\pi g_sN} \big[ d\theta^2+\Delta_0^{-1}(1+{a^2\ov U^2})
\sin^2\theta d\varphi^2 
+ \Delta_0^{-1}\cos^2\theta d\Omega_3^2\big] \bigg]\ ,
}
$$
\Delta_0=1+{a^2\cos^2\theta\ov U^2}\ .
$$
The background asymptotically approaches $AdS_5\times S^5$. The scalar curvature vanishes
identically, but there is a naked singularity at $U=0$, as can be seen
from the  invariant
\eqn\buno{
R_{\mu\nu}R^{\mu\nu}={160\ov \Delta_0^3}\bigg(1+{a^2\ov U^2}+{a^4\ov 4U^4}\cos^2\theta \bigg)^2\ .
}

The metric \atres\ takes a simple form in spheroidal coordinates: 
$$
y_1=\sqrt{r^2+l^2} \sin\theta\cos\varphi\ ,\ \ \ \ 
y_2=\sqrt{r^2+l^2} \sin\theta\sin\varphi\ ,
$$
$$ 
y_3=r\cos\theta\cos\psi_1\ , \ \ \ \ \ 
y_4=r\cos\theta\sin\psi_1\cos\psi_2\ ,\  
$$
$$
y_5=r\cos\theta\sin\psi_1\sin\psi_2\cos\psi_3\ ,\  \ \ \ 
y_6=r\cos\theta\sin\psi_1\sin\psi_2\sin\psi_3\ . 
$$
One finds
\eqn\iis{
ds^2_{\rm IIB}=f_0^{-1/2}\big[ - dx_0^2+dx_1^2+dx^2_2+dx_3^2\big]+f_0^{1/2}
\sum_{m=1}^6 dy_m^2 \ ,
}
where 
$$
f_0=1+{4\pi g_sN{\a'}^2\ov r^2(r^2+l^2\cos^2\theta ) }\ ,
$$
$$
r^2=\ha (\rho^2_1-l^2)+\ha \sqrt{ (\rho_1^2-l^2)^2+4l^2\rho_2^2}\ ,\ \ \ \ 
r^2+l^2\cos^2\theta =\sqrt{ (\rho_1^2-l^2)^2+4l^2\rho_2^2}\ ,
$$
$$
\rho_1^2\equiv  y_1^2+...+y_6^2\ ,\ \ \ \ \ \rho_2^2\equiv y_3^2+....+y_6^2\ .
$$
It is easy to check that $f_0=f_0(y_m)$ is an harmonic function in the space $y_m$. It would be interesting to find an $\N=2 $ four-dimensional
field theory associated with the supersymmetric compactification \iis .

\newsec{Model of QCD}

The backgrounds \once , \docc , generalize
the Witten model \sonce\ by two extra parameters $a_1,a_2$.
Based on the no-hair theorem, in the previous section we have argued
that there are  no other smooth manifolds which can describe
a confining gauge  theory in four dimensions.
It is therefore of interest to explore whether there is a sense in which these compactifications can be
related to non-supersymmetric gauge theories.

Let us first consider the case \once , in which  $a_2=0$. 
It is convenient to work with coordinates $\theta_{1,2}$ which
are $2\pi $-periodic, defined as in \angg ,  with (see \cinco , \btres )
\eqn\auh{
R_0=(2\pi T_H)^{-1}={A\ov 3u_0}\  , \ \ \ \ \ \ \ A\equiv {u_0^4\ov u_H^4-\tre a^4} \ ,
}
where we have introduced the coordinate $u$ by $U=2(\pi N)^{1/2} u$, and redefined 
$a\to 2(\pi N)^{1/2}a $.
By dimensional reduction in the $x_5$ direction, 
we obtain 
$$
ds^2_{\rm IIA}={8\pi \lambda A u^3\ov  3u_0} \Delta ^{1/2} \big[ -dx_0^2+dx_1^2+dx_2^2+dx^2_3\big]+
{8\pi\lambda A^3\ov 27}{u^3\ov u_0^3} \Delta^{1/2} (1-{u_0^6\ov u^6 \Delta }) d\theta_2^2
$$
$$
+\ {8\pi\lambda A\ov 3}{du^2 \Delta ^{1/2} 
\ov u_0u(1-{a^4\ov u^4}-{u_0^6\ov u^6 }) }
$$
\eqn\pocho{
+\ {2\pi\lambda A\ov 3u_0}u\Delta ^{1/2} \bigg(d\theta^2+{\td \Delta\ov \Delta} \sin^2\theta d\varphi^2 
+{1\ov \Delta } \cos^2\theta d\Omega_2^2
 -{4a^2 A u_0^2\ov 3u^4\Delta } \sin^2\theta d\theta_2 d\varphi
\bigg)\ ,
}
\eqn\dilz{
e^{2\phi }={8\pi\ov 27} {A^3\lambda^3 u^3\Delta^{1/2}\ov u_0^3} {1\ov N^2} 
 \ ,\ \ \ \ \ \Delta=1-{a^4\ov u^4} \cos^2\theta \ ,\ \ \ \ \td\Delta=1-{a^4\ov u^4} \ .
}
With this normalization, the metric reduces to eq.~(4.8) of ref.~\edw\ after setting $a=0$
(cf.~eqs.~\socho , \dila ).
The string coupling $e^\phi $ is of order $1/N$, and the metric has become independent of $N$, which is consistent with the expectation that in the large $N$ limit the string spectrum should be independent of $N$. 

The location of the horizon is at 
\eqn\mass{
u_H^6-a^4 u_H^2-u_0^6=0\ ,
}
i.e.
\eqn\uuhh{
u_H^2={a^4\ov \gamma u_0^2}+ \tre \gamma u_0^2\ ,\ \ \ \ \ 
\gamma=3\bigg(\ha+\ha \sqrt{1-{4\ov 27}\big({a\ov u_0}\big)^{12} } 
\bigg)^{1/3}\ .
}
For all  $a^4, u_0^6>0$, $u_H^2$ is a positive real number, $u_H>a>0$.

The spectrum of the Laplacian corresponding to the space of this model, eq.~\once , 
has a mass gap . This is characteristic of
smooth geometries with a cutoff at some $u=u_0$ that approach $AdS$ at infinity: there are no oscillatory
solutions to the Laplace equation at infinity, so the spectrum of normalizable eigenfunctions that are regular at $u=u_0$ must therefore be discrete \edw .
The isometry group of the internal space is $SO(3)\times U(1)\times U(1)$.
The supergravity fields on this space can be classified in terms of 
representations of $SO(3)$, given by the standard spherical harmonics.
Glueballs of QCD are presumably related to $SO(3)$ singlets, with vanishing $U(1)\times U(1)$ charges. 

Let us consider the following Laplace operators corresponding to the background 
\pocho ,~\dilz : 
$$
\nabla^2\equiv {1\ov \sqrt{g} }\del_\mu \sqrt{g}
g^{\mu\nu}\del_\nu \ ,\ \ \ \ 
\hat \nabla^2\equiv {e^{2\phi} \ov \sqrt{g} }\del_\mu \sqrt{g}e^{-2\phi }
g^{\mu\nu}\del_\nu \ ,\ 
$$
$$
\sqrt{g}= C\  u^9 \Delta \cos^2\theta\sin\theta \sin\psi_1 \ ,\ \ \ \ \  C={1\ov 2\pi \lambda }
\left( {4\pi\lambda A\ov 3 u_0 }\right)^6 \ .
$$
It is easy to see that there are no solutions to the equation $\nabla^2 \Phi =0$ of the form
$\Phi=\chi (u) e^{ik.x} $, one needs $\Phi=\chi (u, \theta ) e^{ik.x} $.
Interestingly,  there are however $\theta $-independent solutions 
$\Phi=\chi (u) e^{ik.x} $
to the equation $\hat \nabla^2 \Phi =0$ . One obtains
\eqn\glus{
{1\ov u^3} \del_u u (u^6-a^4 u^2-u_0^6) \del_u \chi (u)= -M^2 \chi (u)\ ,\ \ \ \ \ M^2=-k^2\ .
}
This equation can be treated, for instance, by using the WKB method.
Glueballs are associated with the massless scalar that couples to the
operator $F_{\mu\nu}F^{\mu\nu}$. This obeys the Laplace equation in the Einstein frame
(cf.~ref.~\horstro ), which in the string frame  corresponds to
$\hat \nabla^2\Phi=0$.
Therefore glueball masses should be determined by eq.~\glus .

An estimate of the glueball masses can be given from the Yang-Mills string tension.
The string tension can be determined by computing the Wilson loop as in \mwils . We look for string configurations that minimize the Nambu-Goto action.
The new feature is that now the metric components depend on $\cos^2\theta $. Let us first consider configurations with constant $\theta $. This is a solution provided
$$
{\delta \cos^2\theta \ov \delta \theta }=0\ ,
$$
i.e. $\theta=0, {\pi\ov 2},\pi $. The string tension is then given by 
${1\ov 2\pi }$ times the coefficient of $\sum dx_i^2 $, evaluated at the
horizon $u=u_H$, that is (cf. eq.~\sst )
\eqn\tay{
T_{\rm YM}(\theta=0,\pi )={4\ov 3} \lambda A {u_H^3\ov u_0} \sqrt{1-{a^4\ov
u_H^4} }\ ,
}
\eqn\tayy{
T_{\rm YM}(\theta=\ha \pi )={4\ov 3} \lambda A {u_H^3\ov u_0} \ .
}
The string configuration with minimal area is thus at $\theta=0,\pi $, with
the tension given by \tay . 
For configurations with non-constant $\theta =\theta (\s )$ one needs to find the tension by explicitly computing the Nambu-Goto action on the solution
$U(\s ), \theta (\s )$. The system of equations is highly non-linear, but we expect that the absolute minimum is at $\theta =0,\pi $, for which $\Delta^{1/2} $ (and hence $T_{\rm YM}\propto \Delta^{1/2}$) takes the 
minimum value that it can have.
Using \mass , the tension \tay \ becomes
\eqn\tayz{
T_{\rm YM}={4\ov 3} \lambda A u_0^2={4\ov 3} \lambda {u_0^6\ov u_H^4-\tre a^4}
\ .
}
Thus masses of glueballs associated with excitations of the string
with tension $T_{\rm YM}$ should be of the order 
$\big(T_{\rm YM} \big)^{1/2}$.
The  masses of the  supergravity glueballs 
--~associated with the zero modes of the string~-- are determined from the Laplace equation.
Since $\lambda $ appears only as an overall factor in the metric \pocho ,
these masses are independent of $\lambda $ (they may however receive 
corrections suppressed by factors $1/\lambda $), and they are
of the form $M_{\rm glueballs}=u_0 f(a/u_0)$.
The mass scale should be of order
\eqn\gggs{
M_{\rm glueballs}\cong \Lambda_{\rm QCD}\cong {u_0^3\ov 
\sqrt{u_H^4-\tre a^4} }\ .
}
On the other hand, Kaluza-Klein states associated with the circle parametrized by $ \theta _2$
(of the form $\Phi=\chi (u, \theta ) e^{ik.x} e^{in\theta _2}$)  will have masses
\eqn\kzkz{
M_{\rm KK}\cong R_0^{-1}={3\ov u_0^3}\big( u_H^4-\tre a^4 \big)\ .
}
This means
\eqn\ayay{
{M_{\rm glueballs}\ov M_{\rm KK} }\cong A^{3/2}= {u_0^6\ov (u_H^4-\tre a^4)^{3/2}}\ .
}
This is a function of $a/u_0$ (see \uuhh ), and the behavior at small and
large $a/u_0$ is as follows:
\eqn\kakk{
M_{\rm glueballs}\cong  M_{\rm KK}  +O(a^4)\ ,\ \ \ \ \ \ \   u_0\gg a\ ,
}
\eqn\kaka{
{M_{\rm glueballs}\ov  M_{\rm KK}} \cong {u_0^6\ov a^6}\ll 1  \ ,\ \ \ \ \  \ \ \ \ \ \ u_0\ll a\ .
}
For the glueballs related to  string excitations, the ratio of masses will be
\eqn\kkk{
{\big(T_{\rm YM} \big)^{1/2}\ov M_{\rm KK} }
\cong \sqrt{\lambda } \ {u_0^6\ov a^6 }\ ,\ \ \ \ \ u_0\ll a\ .
}
In order for the  Yang-Mills tension $T_{YM}\cong \lambda \ u_0^6/a^4$ to remain finite at large $a$ (and fixed $u_0$) one has to take $\lambda \sim a^4/u_0^4 $.
Thus for $a\gg u_0$ the low-energy theory will be effectively $3+1$ dimensional
with a finite value for the string tension, and hence 
a finite value for glueball masses. In this sense $a$ (which has dimension of mass)
acts as a momentum cutoff.
For  $u_0\gg a$,  one recovers the relation \kakk\  of the Witten model.

Having a regime $u_0\ll a$ where $M_{\rm glueballs}\ll M_{\rm KK}$, the next question
is where the supergravity approximation applies.
For this it is necessary that curvature invariants are small in the whole space. The maximum
curvature is attained at $u=u_H$ and at the poles, $\theta=0,\pi $. For $u_0\ll a$ one has 
$\a' R \sim u_0/(u\lambda A\Delta^{1/6})$. Using eqs.~\auh ,~\dilz \ and \mass\ we find
$$
\a' R< O({a^4\ov\lambda u_0^4}) \ ,\ \ \ \ \ \ \ u_0\ll a \ .
$$
Therefore the regime where  we can use supergravity and at the same time decouple the Kaluza-Klein particles with $U(1)$ charge is  large $\lambda $ and large $a/u_0$, with $\lambda \gg a^4/u_0^4 $. [In general, for any given $a$ and $u_0$ one can pick a sufficiently large $\lambda $ such that all curvature invariants are small].
This is precisely the region where the string excitations of mass $\big(T_{\rm YM} \big)^{1/2}$ become heavy.
The supergravity approximation only incorporates particles which 
have spin $\leq 2 $. In order to incorporate 
the full Regge trajectories of spin $J$-glueballs with squared masses of 
order $T_{\rm YM} |J|$, one needs to use string theory.\foot{We thank J.~Maldacena for comments on this point.}
Although  the string spectrum in this background is a function of $\lambda ,\ u_0$ and $a$, which can be quite complicated, it is nonetheless possible to anticipate 
a hierarchy of scales as follows.
Equation \tayz\ indicates that the masses of excitations of the string increase with $\lambda $: for $\lambda \gg 1$ there will be a gap between the lowest glueball states of supergravity   and the glueballs represented by string excitations. The analysis in eqs.~\kakk -~\kkk\ shows that at $a/u_0\gg 1$
there will be another gap separating the scale of the string excitations from the scale of Kaluza-Klein particles with $U(1)$ charge.


The relation between confinement and monopole condensation was recently discussed in this context by Gross and Ooguri \groo . The magnetic monopole is represented by  a D2 brane wrapped around the $\theta_2$ circle that ends on a D4 brane. The potential between a monopole $m$ and an anti-monopole $\bar m$   can then be computed by considering a D2 brane bounded by $S^1$ times the trajectories of $m$ and $\bar m$.
Let us consider a configuration with $\theta=0$ (or $\theta=\pi $), with the monopoles travelling along $x_1$
 and separated in the $x_2$ direction by a distance $L$.
{}Using eq.~\pocho\ we find for the D2 brane action
$$
E_{m\bar m}=\int _0^L d\theta_2 dx_2 \ e^{-\phi }\sqrt{g_{\theta_2\theta_2}g_{11}g_{22} }
$$
\eqn\emmp{
={1\ov 2g_sT_H} \int _0^L dx_2\sqrt{ \left( {d\bar u\ov dx_2}\right)^2 
+{1\ov g_s N} \big( \bar u^3-\bar a^2\bar u-\bar u_0^3\big) }\ ,
}
where $\bar u=4g_sN u^2 ,\ \bar a=4g_sN a^2$. When $\bar a=0$, this reduces 
to eq.~(12) of \groo .
By minimizing the action \emmp , one can see that the phenomenom of complete screening
of the magnetic monopole observed in \groo\ persists for any value of $\bar a,\bar u_0>0$.
This happens because when $L$ goes beyond some critical distance $L_{\rm crit} $ there is no connected
$D2$ brane configuration minimizing the action.
For a connected minimal surface, the distance $L$ between $m$ and $\bar m$ can be expressed in terms of the minimum value of $\bar u=\bar u_{\rm m}$ 
reached by the D2 brane as follows
\eqn\nomi{
L={2\sqrt{g_sN}\ov \bar u_{\rm m}^2}\sqrt{\bar u_{\rm m}^3-\bar a^2 \bar u_{\rm m}-\bar u_0^3}
\int_1^\infty {dy \ov  \sqrt{
\left(y^3-{\bar a^2\ov \bar u_{\rm m}^2}y -{\bar u_0^3\ov \bar u_{\rm m}^3}\right)
\left(y^3-1-\bar a^2 (y-1)\right)} }
\ .
}
This integral can be evaluated numerically as a function of $\bar u_{\rm m}$, $\bar u_H<\bar u_{\rm m}<\infty $, showing that there is indeed a maximum distance $L_{\rm crit} $ for any
real $\bar a, \bar u_0 >0$. As a result, the potential between monopoles becomes constant for $L>L_{\rm crit}$.

Finally, let us consider the case \docc , where there are two non-vanishing components of the
``angular momentum" parameters $a_1, a_2$ . 
The dilaton is now given by \dilz\ with
$$
A^{-1}={1\ov u_0^4}\bigg[u_H^4-\tre (a_1^4+a_2^4)-{a_1^4a_2^4\ov 3u_H^4}
\bigg]\ ,
$$
\eqn\abcd{
\Delta= 1-{a_1^4\ov u^4} \cos^2\theta -
{a_2^4\ov u^4}\big( \cos^2\theta \cos^2\psi_1 +\sin^2\theta\big)
+{a^4_1a^4_2\ov u^8}\cos^2\theta \cos^2\psi_1 \ ,
}
and $u_H$ is the largest solution of the equation
$$
\big(1- {a_1^4\ov u^4_H}\big)\big(1- {a_2^4\ov u^4_H}\big)={u_0^6\ov u_H^6}\ .
$$
Note that  for any $u_0^6> 0$ one has $u_H>a_1,a_2$.

 To compute 
the Wilson loop we consider a configuration with constant $\theta$,
$\psi_1$, which mimimizes the Nambu-Goto action provided
${\delta \cos^2\theta \ov \delta \theta }=0\ $, 
${\delta \cos^2\psi_1 \ov \delta \psi_1 }=0\ $, i.e.
$\theta, \psi_{1}=0,\ \ha \pi,\ \pi$. 
The string tension is thus given by
\eqn\ttym{
T_{\rm YM}={4\lambda R_0 }  u_H^3 \Delta_H^{1/2}\ ,\ \ \ \ \ R_0={A\ov 3u_0}
}
The tension takes the minimum value when $\theta$ and $\psi_{1}$ are equal to $0$
or $\pi $,
so that
\eqn\qqss{
\Delta_H= \big(1- {a_1^4\ov u^4_H}\big)\big(1- {a_2^4\ov u^4_H}\big)={u_0^6\ov u_H^6}
 \ ,
}
which leads to
$$
T_{\rm YM}={4\lambda R_0 u_0^3 }\ .
$$
Thus
$$
{M_{\rm glueballs}\ov  M_{\rm KK}} \cong (R_0u_0)^{3/2} \cong A^{3/2}\ .
$$
As in the previous $a_2=0$ case, a $3+1$ dimensional theory with a 
finite scale $\Lambda_{QCD}$ 
seems to arise in some regime with large $a_1$ or large $a_2 $.

The mass scale corresponding to 
non-singlet representations of $SO(3)$ cannot be estimated in a simple way. 
To establish what are the precise mass scales of the different particles of the spectrum a thorough analysis is obviously needed. This may require a
numerical treatment of the Laplace equation as the one carried out in \refs{\oogu -\ogkk }.

\bigskip
\bigskip

\noindent{\bf Acknowledgements}
\medskip
The author wishes to thank J.~Maldacena for useful comments.
He would also like to thank the Theory Division of CERN, where part of this work was carried out,
for the kind hospitality.
This work is supported by the European
Commission TMR programme grant  ERBFMBI-CT96-0982.

\vfill\eject
\listrefs

\bye